\documentclass{article}
\usepackage{ijcai22}
\usepackage{fancyhdr}

\fancypagestyle{firstpage}{
    \fancyhead[L]{}    
    \fancyhead[R]{}  
    \fancyhead[C]{This paper was presented at the 35th International Workshop on Qualitative Reasoning, IJCAI\_ECAI Conference, July 23-29, 2022, Messe Wien, Vienna, Austria.}}

\usepackage{times}
\usepackage{soul}
\usepackage{url}
\usepackage[hidelinks]{hyperref}
\usepackage[utf8]{inputenc}
\usepackage[small]{caption}
\usepackage{graphicx}
\usepackage{amsmath}
\usepackage{amsthm}
\usepackage{booktabs}
\usepackage{algorithm}
\usepackage{algorithmic}
\usepackage{lipsum}
\title{Contextualizing Large-Scale Domain Knowledge for \\Conceptual Modeling and Simulation}

\author{
Sungeun An$^1$
\and
Spencer Rugaber$^1$\and
Jennifer Hammock$^2$\And
Ashok K. Goel$^1$
\affiliations
$^1$School of Interactive Computing, Georgia Institute of Technology\\
$^2$National Museum of Natural History, Smithsonian Institution
\emails
sungeun.an@gatech.edu
}

\begin{document}

\maketitle

\begin{abstract}
    We present an interactive modeling tool, VERA, that scaffolds the acquisition of domain knowledge involved in conceptual modeling and agent-based simulations. We describe the knowledge engineering process of contextualizing large-scale domain knowledge. Specifically, we use the ontology of biotic interactions in Global Biotic Interactions, and the trait data of species in Encyclopedia of Life to facilitate the model construction. Learners can use VERA to construct qualitative conceptual models of ecological phenomena, run them as quantitative simulations, and review their predictions.
\end{abstract}
\thispagestyle{firstpage}

\section{Introduction}
Modeling is central to human cognition and scientific reasoning \cite{schwarz2005metamodeling}. Qualitative representations in the form of conceptual models capture the components and mechanisms that explain the observed phenomenon \cite{forbus1984qualitative}\cite{schwarz2005metamodeling}, which helps learners externalize, share, and simulate the system the model represents. Simulation models are executable with specific values for the system’s input variables, enabling determination of the temporal evolution of the values of the system’s output variables \cite{white1990causal}\cite{de1998scientific}. 

The importance of learning how to construct, use, evaluate, and revise models has been advocated by many researchers \cite{frederiksen2002conceptualizing}. In particular, with the importance of addressing environmental problems, there is growing interest in empowering novice learners with little scientific training or expertise to engage in modeling activities to increase their understanding of the local environments and to learn through various experiments. They are also interested in capturing and simulating conceptual knowledge \cite{huang2017scientific}.  

However, many studies have revealed that modeling is a difficult process for novice learners and hampered by lack of domain knowledge \cite{hogan2001cognitive}\cite{sins2005difficult}\cite{vanlehn2013model}. Ecological knowledge is inherently heterogenous, including both qualitative and quantitative aspects \cite{salles2003qualitative}. The qualitative aspect of ecological knowledge includes definitions of components, a set of variables associated with them, and causal relations between the variables associated with the different components \cite{liem2013supporting}. For example, learners should be able to answer some questions, such as ``How much energy will an alligator acquire from eating a muskrat?'' and ``What are the relationships between body mass and respiratory rate?'' Without the domain’s necessary knowledge as a pre-requisite, the learner will not be able to benefit from using and creating models because there is no knowledge to differentiate and integrate \cite{tennyson2002improving}. 


Of course, large amounts of knowledge about many domains are now readily accessible on the internet. However, much of this general-purpose knowledge is not particular to any specific task and thus difficult to comprehend by many learners. The research question then becomes can we provide access to large-scale domain knowledge in a comprehensible manner? Our research hypothesis is that access to large-scale domain knowledge should be contextualized and that contextualized acquisition of this knowledge may help learners achieve deeper understanding about the domain and generate richer models.

In this paper, we describe the knowledge engineering process of contextualizing large-scale domain knowledge in an online modeling called VERA. The Virtual Experimentation Research Assistant (VERA; \url{vera.cc.gatech.edu}; \cite{an2020scientific}) supports ecological modeling using large-scale domain knowledge through Smithsonian's Encyclopedia of Life [EOL; \url{eol.org}; \cite{parr2016traitbank}] and Global Biotic Interaction [GloBI; \url{globalbioticinteractions.org}; \cite{poelen2014global}]. Specifically, we used the ontology of biotic interactions in GloBI and the trait data of species in EOL to facilitate the model construction. Learners can use VERA to construct qualitative conceptual models of ecological phenomena, run them as quantitative simulations, and review their predictions.

\section{Related Work}
Much cognitive systems research has explored interactive tools for qualitative modeling and qualitative simulation and their use for promoting science education and research \cite{bredeweg2003qualitative}\cite{forbus2005vmodel}\cite{leelawong2008designing}\cite{bredeweg2013dynalearn}. VERA has a close similarity to Forbus et al.'s Vmodel [2005], Bredeweg et al.’s Garp-3 system [2009], and DynaLearn [2013] in that it allows the user to first create qualitative models of ecological phenomena and then qualitatively simulate them \cite{forbus2005vmodel}\cite{bredeweg2009garp3}\cite{bredeweg2013dynalearn}. In contrast to these qualitative modeling tools, VERA uses Component-Mechanism-Phenomena language to assist the process of authoring conceptual models. VERA also uses a translator between the conceptual models and the off-the-shelf NetLogo engine for directly spawning the agent-based simulations from the conceptual models. 


VERA is built on the previous work that used Component-Mechanism-Phenomenon (or CMP) causal models \cite{joyner2011evolution} that arose from Structure-Behavior-Function models \cite{goel2009structure}. In CMP, Components represent the Structure of the system, emphasizing the physical pieces of an ecological system and their properties. Mechanism represents the Behavior of the system, a sequence of developments linked together in a causal chain that results in some observable Phenomenon. Phenomenon represents what was formerly described as the Function of the system; the Phenomenon is the initial observable event that the investigator aims to explain. VERA is also built on the previous work that used an off-the-shelf agent-based simulation system called NetLogo \cite{joyner2014mila} because agent-based simulations are especially well suited for ecological modeling \cite{wilensky2006thinking}.


Although the focus of this paper is on the knowledge engineering in VERA, we should note that earlier work has demonstrated VERA's usefulness for enabling learning of domain knowledge \cite{an2018vera}. In particular, we found that learners who used the full cycle of model construction, evaluation, and revision created models of higher quality than learners who only constructed conceptual models or only observed the simulations \cite{an2022understanding}.

\section{VERA}
VERA (Virtual Experimentation Research Assistant) is an online learning environment that enables learners to construct conceptual models of ecological systems and run agent-based simulations of these models. This allows learners to explore ecological systems and perform ``what-if'' experiments to either explain an existing ecological system or attempt to predict the outcome of future changes to one.

From a human-learning perspective, the key steps in the overall procedure for learning in VERA are as follows. A learner often starts with the identification of an atypical or abnormal phenomenon (e.g., overpopulation). The learner may then develop multiple hypotheses for explaining the phenomenon. The VERA's interface for CMP conceptual modeling coupled with the contextualized domain knowledge helps the learner to express and elaborate his or her hypotheses. The learner may then evaluate his or her hypotheses through the agent-based simulations, which may lead to a revision of his or her hypothesis.

\begin{figure}[h]
\centering
\includegraphics[width=0.45\textwidth]{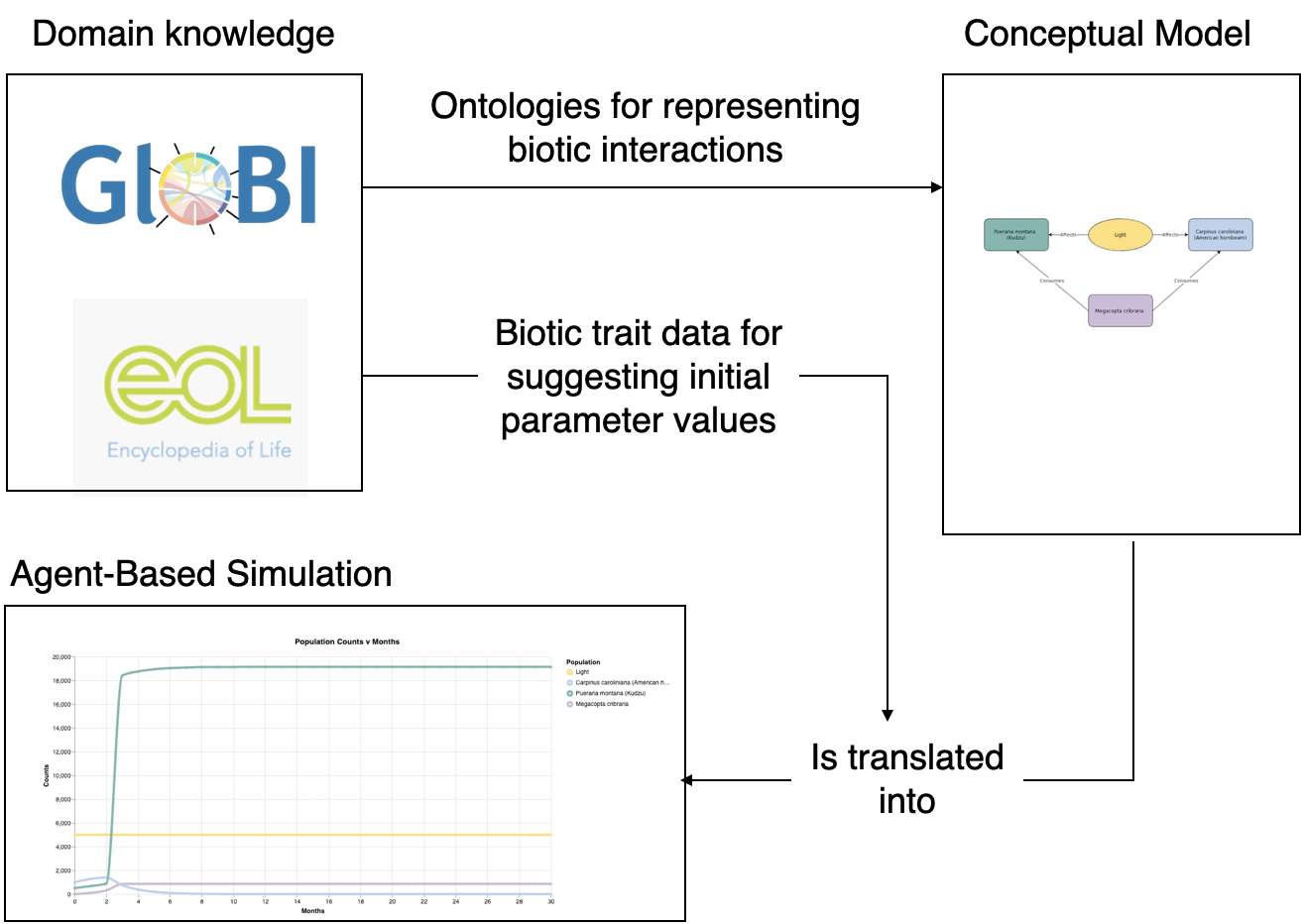}
\caption{Schematic Overview of Using Domain Knowledge for Conceptual Modeling and Simulation.}

\label{fig1}
\end{figure}

This paper focuses on describe the knowledge engineering process of contextualizing large-scale domain knowledge in VERA. As shown in Figure \ref{fig1}, VERA uses domain knowledge in two ways. First, VERA's taxonomy of interactions among biotic components in a conceptual model is based on the ontology of the interactions used by Global Biotic Interactions (GloBI). Second, VERA uses trait data from Encyclopedia of Life to suggest the initial parameter values and define the relationships between variables. 

This section introduces the design of our system including conceptual models using CMP modeling language (Section \ref{conceptual_model}) and an AI compiler that translates the conceptual models into agent-based simulation (Section \ref{translating}). Section \ref{contextualize_knowledge} provides how VERA retrieves and contextualize domain knowledge.

\subsection{Conceptual Models}\label{conceptual_model}
The declarative conceptual models of an ecological phenomenon in VERA are based on the Component-Mechanism-Phenomenon (CMP) models. Components in VERA can be either biotic or abiotic, and each component has a set of variables associated with it, thirteen for biotic and three for abiotic components to represent the structure of the system. For example, biotic components are defined by their \textit{lifespan, reproductive maturity, reproductive interval, offspring count, starting population, minimum population, body mass, carbon biomass, respiratory rate, photosynthesis rate, assimilation efficiency, move direction, and move velocity}. Abiotic substances are defined by their \textit{amount, minimum amount, and growth rate}. These properties were selected and adapted to represent ecologically relevant attributes and primitives for ecological modeling (see Table \ref{tab1}).

\begin{table*}[t]
    \centering
    \begin{tabular}{p{0.1\linewidth}  p{0.1\linewidth}  p{0.3\linewidth} p{0.3\linewidth}}  
    
        \toprule
        Component  &  Property & Description & EOL Trait Data \\
        \midrule
        Biotic     & Lifespan  & Average lifespan of organisms in this population in months. & “life span” or “total life span”  \\
                   & Reproductive Maturity & Age when organisms in this population are able to begin reproduction in month. & “age at first birth”,“age at first reproduc-tion”,“age at maturity”,“onset of fertility”,“egg laying begins” \\
                   &Reproductive Interval&Frequency with which organisms in this population are able to reproduce in months.& “inter-birth interval” \\
                   & Offspring Count & Average number of offspring per spawning individual for a reproduction cycle.&“offspring” “litters per year”\\
                   &Body Mass&Average body mass per organism. &“body mass” If it is not available, attempt to estimate it based on taxonomic ancestry traits (“body length”)\\
                   &Carbon Biomass&Average carbon biomass in an individual organism. &“carbon biomass” Attempt to estimate it based on taxonomic ancestry traits (“plant height”, “body mass”)\\
                   &Respiratory Rate&Average basal metabolic rate, measured as respiration (loss) of carbon biomass.&“respiratory rate” If it is not available, attempt to estimate it based on taxonomic ancestry traits (“basal metabolic rate”, “body mass”)\\
                   &Photosynthesis Rate&Average addition of carbon biomass from photosynthesis for a square meter of density-based populations (kg/month).&“photosynthetic rate” If it is not available, attempt to estimate it based on taxonomic ancestry traits (“net carbon fixation rate”)\\
                   &Assimilation Efficiency&Efficiency of assimilating carbon biomass via consumption (0.0 - 1.0).&Attempt to estimate it based on taxonomic ancestry traits. \\
                   
        \bottomrule
    \end{tabular}
    \caption{The Selected Properties of Biotic Components and their Trait Data for Conceptual Modeling.}
    \label{tab1}
\end{table*}

To describe casual relationships among components in an ecological system, we used the ontology of the interactions from a digital library called Global Biotic Interactions (GloBI) \cite{poelen2014global}. GloBI provides open access to finding species interaction data (e.g., predator-prey, pollinator-plant, pathogen-host, parasite-host) by combining existing open datasets using open-source software. GloBI provides 22 possible interaction types between species. Among 22 interaction types, we reduced to five interaction types by integrating redundant interactions and extracting ``primitive'' interactions between components that give rise to the behavior of the system as a whole (see Table \ref{tab2}). Examples of such primitive component–component interactions include \textit{consumes} (one biotic organism consuming another), \textit{produces} (a biotic organism producing an abiotic substance), and \textit{destroys} (an abiotic substance harming a biotic organism). 

\begin{figure*}[h]
\centering
\includegraphics[width=0.8\textwidth]{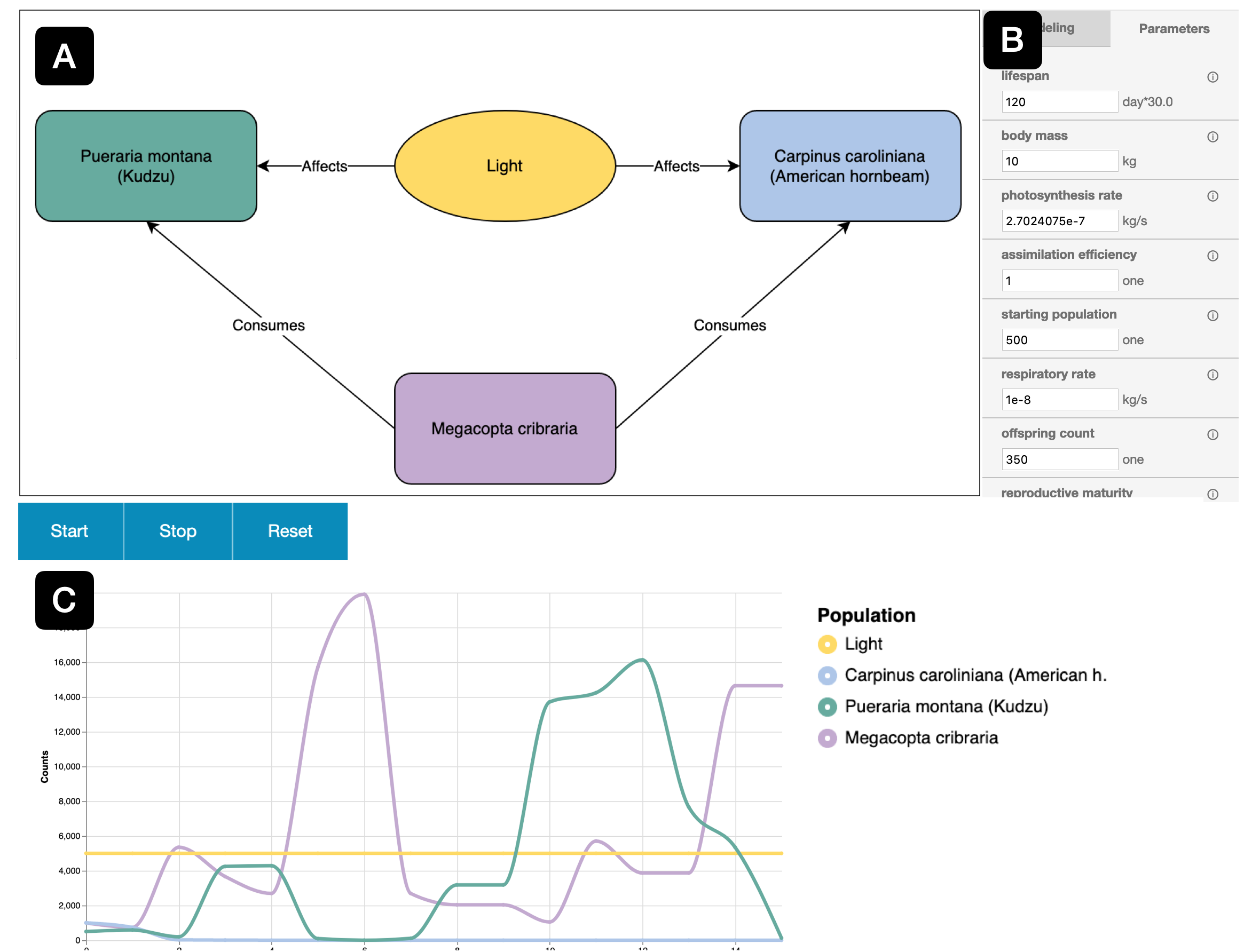}
\caption{The VERA system. (A) Conceptual Model. (B) Simulation parameters. (C) The Simulation Output Graph – x axis: Time (months); y axis: Population.}
\label{fig2}
\end{figure*}

To instantiate the principles of CMP and better facilitate modeling of ecological phenomena, VERA provides a visual interface for Component-Mechanism-Phenomenon language with a well-defined semantics to represent conceptual models clearly. Figure \ref{fig2}(A) shows a screenshot of the model canvas in VERA where a learner can build a conceptual model by adding biotic (rectangular), abiotic (ellipse), and relationships among them. The learner can draw a directed relationship between two components and choose a pre-dfined relationship type from the drop-down menu. The right-side panel in Figure \ref{fig2}(B) is the modeling and simulation parameter editor. The simulation parameters of each component can affect its simulation behavior. 

Figure \ref{fig2}(A) illustrates a causal conceptual model of kudzu (\textit{Pueraria Montana}), kudzu bug (\textit{Megacopta Cribraria}), and American hornbeam (\textit{Carpinus Caroliniana}) in the Southern United State. In this model, there are four components: light (abiotic), kudzu (biotic), American hornbeam (biotic), and kudzu bug (biotic). Regarding mechanisms (relationships), kudzu bug consumes both kudzu and American hornbeam, and light affects both kudzu and American hornbeam. The direction of the arrow between the variables of two components indicates the direction of causal influence. For example, the arrow from kudzu bug to kudzu indicates that the population of kudzu bug consumes the population of kudzu.

A Phenomenon is an observation of the system of interest. For example, the mechanism illustrated in Figure \ref{fig2}(A) can exhibit three different phenomena depending on the manipulation of the kudzu bug population. First, when kudzu bug is low, kudzu grows fast and outcompetes American hornbeam for the shared resource of light, and American hornbeam does not survive the competition with kudzu. Second, when the population of the kudzu bug is adequate (medium), the kudzu population is controlled while American hornbeam also survives (as shown in Figure \ref{fig2}(C)). Lastly, when the population of the kudzu bug is high, the Kudzu and the American hornbeam population both die off due to the large kudzu bug population.

\subsection{Contextualize Domain Knowledge}\label{contextualize_knowledge}
\begin{figure}[h]
\centering
\includegraphics[width=0.45\textwidth]{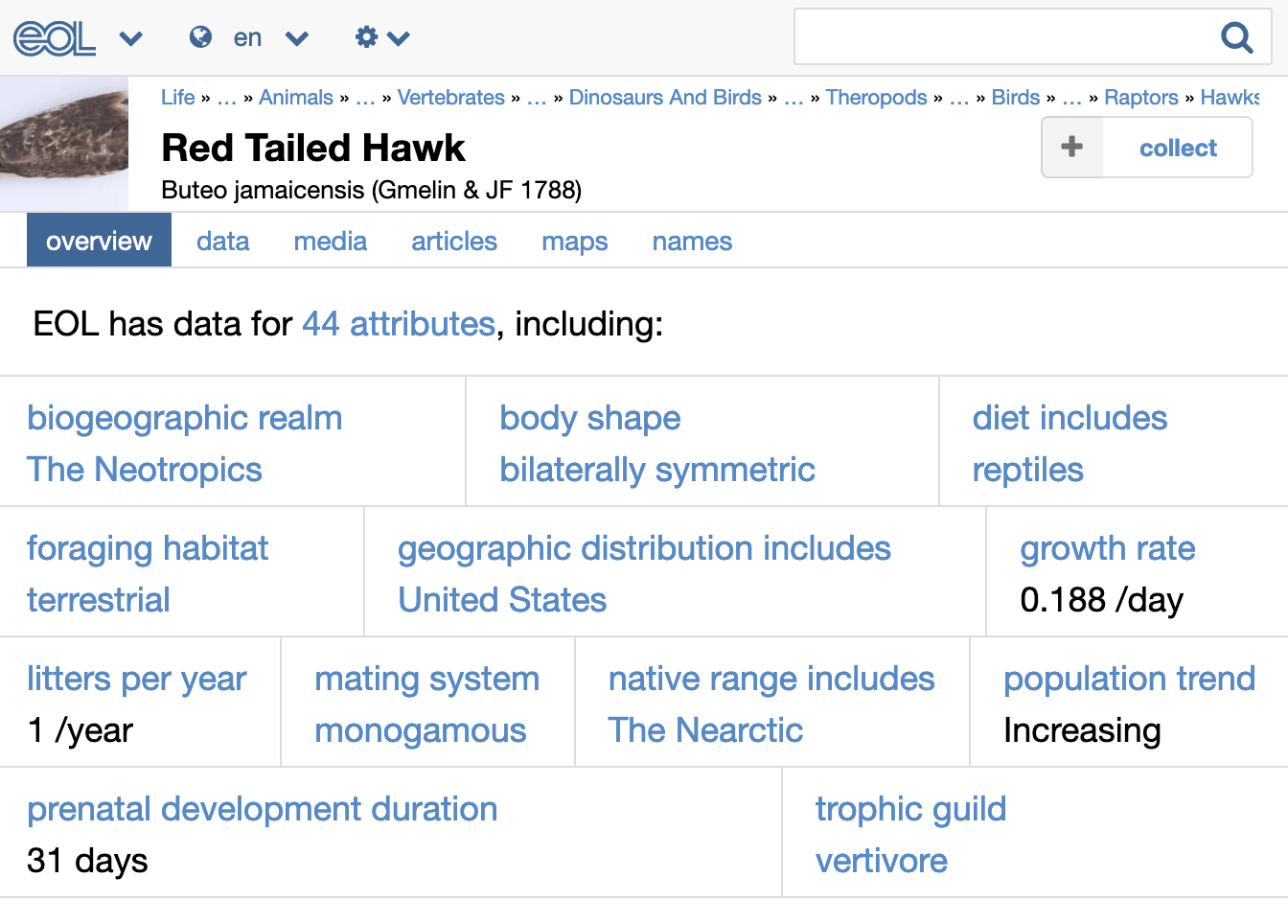}
\caption{Screenshot of EOL trait data of Red Tailed Hawk}
\label{fig3}
\end{figure}

We use EOL's the trait data to suggest initial parameter values for species of interest (see Table \ref{tab1}). EOL is the world’s largest aggregated and curated database of species data with almost two million species and eleven million attribute records in the biological domain \cite{parr2016traitbank}. Over 10 million trait data records are available for 1.7 million taxa. Figure \ref{fig3} shows a screenshot of trait data of Red Tailed Hawk in EOL. Trait-Bank aggregates trait records from many sources, and for a given trait, multiple records may exist from different studies under a variety of conditions. To support large-scale queries, EOL provides API services that allow queries for on-demand JSON output about EOL taxa, ecological interactions, and organism attributes in the Cypher language. 

VERA derives simulation parameters from EOL traits if possible, otherwise attempts to estimate them based on trophical ancestor and its body mass. If the trait data is available in EOL, values are directly derived from the EOL trait data. For example, in Table \ref{tab1}, \textit{lifespan} is retrieved from existing EOL traits such as \textit{life span} or \textit{total life span} (In case of redundancy, the average value is used). \textit{Reproductive maturity} is retrieved from traits such as \textit{age at first birth}, \textit{age at first reproduction}, \textit{age at maturity}, \textit{onset of fertility}, \textit{egg laying begins}. However, if the necessary trait data is not available in EOL, VERA attempts to estimate the value with the information of the species' ancestor. For example, if \textit{carbon biomass} is not available in EOL, it is calculated using its \textit{body mass}: $C_{reptilia} (.122) \times bodymass$ (in case its ancestor is Reptilia in freshwater); $C_{mammalia}  (.16) \times bodymass$ (in case of Mammalia). If its taxonomic ancestry traits are not available, VERA provides reasonable default values for ecological plausibility (e.g., $C_{default} (.1) \times bodymass$).

\begin{table*}[t]
    \centering
    \begin{tabular}{p{0.15\linewidth} p{0.15\linewidth} p{0.3\linewidth}  p{0.3\linewidth}}  
    
        \toprule
        Relationship  & Property & Description & GloBI Interactions \\
        \midrule
        X Consumes Y &Consumption rate, interaction probability& When X interacts with Y, it will partially or wholly consume Y, with carbon transfer to X from Y. & “eat”, “get eaten by”, “preys on”, “get preyed on by”\\
        X Destroys Y & Destruction rate, interaction probability&When X interacts with Y, it partially or wholly destroys a simulation entity of type Y with no carbon transfer to X.   & “kill”, “is killed by”, “parasi-tize”, “get parasitized by”, “get infected by” \\
        X Produces Y &Production rate& X will produce Y with some stochastic timing and amount. & “visits flowers of”, “flowers visited by”, “pollinate”, “get pollinated by”, “spread”, “get spread by.” \\ 
        X Affects Y &Growth rate, interaction probability& This is a generic growth modifier that allows for growth rates (negative or positive) to modify Y when X interacts with it, where none of the above relationships apply. & “interacts with” (+, -), “related to” (+, -), “parasitize” (-), “get parasitized by” (-), “hosts”, “get hosted by.” \\
        X Becomes Y on Death  &Percent body mass& When X expires, it produces Y. & - \\
        \bottomrule
    \end{tabular}
    \caption{The Taxonomy of Interactions among Components.}
    \label{tab2}
\end{table*}

\begin{figure*}[h]
\centering
\includegraphics[width=0.8\textwidth]{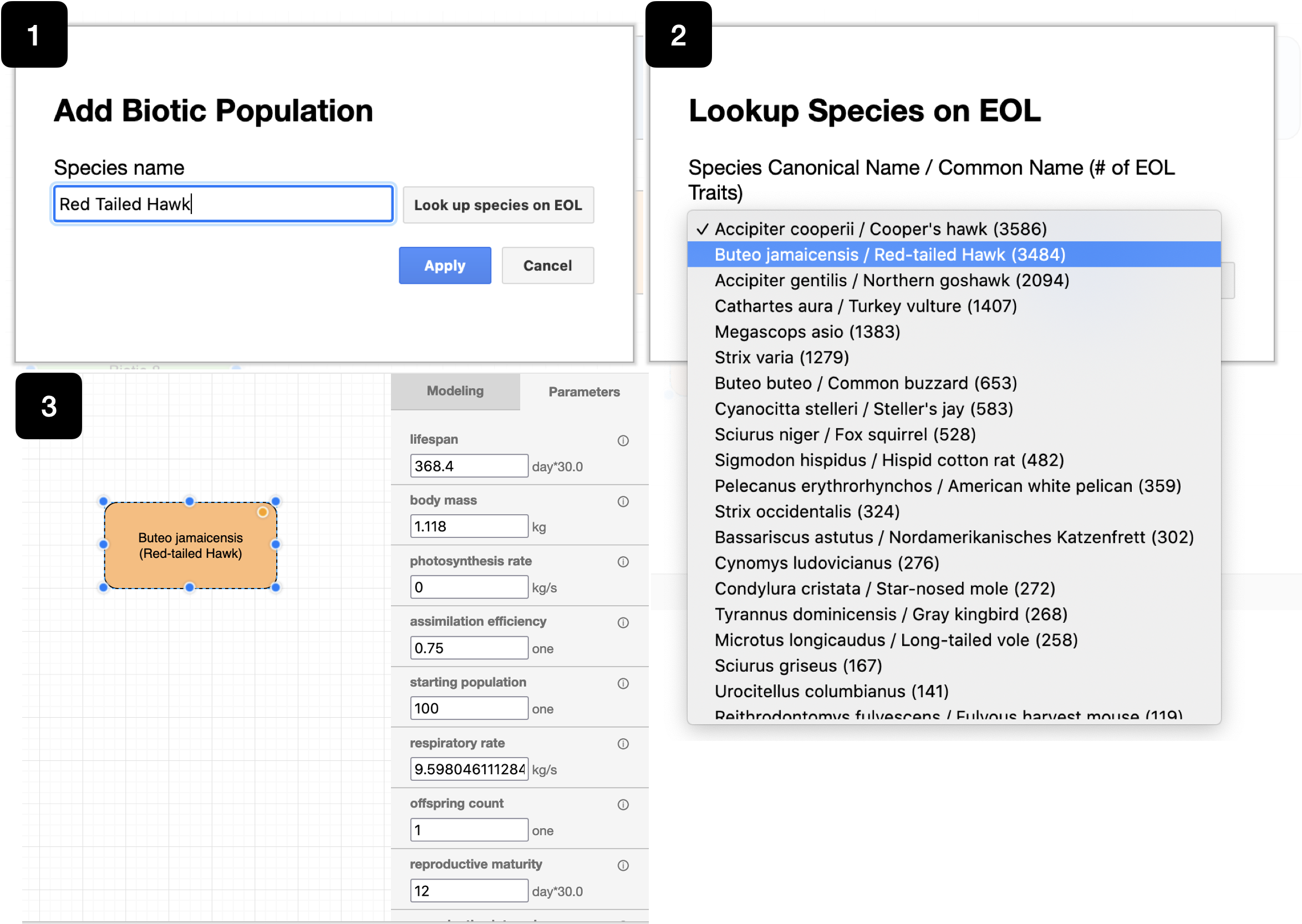}
\caption{Using EOL TraitBank Data to Set Up Simulation Values.}
\label{fig4}
\end{figure*}

VERA provides the contextualized domain knowledge via ``Lookup EOL'' feature. Figure \ref{fig4} shows the process of adding a biotic component via Lookup EOL. (1) The learner clicks on ``LookUp EOL'' button in VERA and queries species name (either scientific or common names). (2) Then the system returns a list of species names that matches the input via EOL Search API. (3) Then, the learner selects one species from the list, and the system calls EOL TraitBank API for retrieving specific traits of the species. Our inference engine uses the retrieved traits to preset the simulation parameters.  

\subsection{Translating Conceptual Models into Simulations}\label{translating}
Following our earlier work on the MILA-S system \cite{joyner2011evolution}\cite{joyner2014mila}\cite{goel2016preparing}, VERA uses an artificial intelligence compiler to automatically translate the patterns in the conceptual models into the primitives of agent-based simulation of NetLogo. In other words, VERA automatically generates the simulations directly from learner's casual model. After constructing a CMP conceptual model, the learner uses the suggested initial parameter values by EOL for simulation generation (as described in Section \ref{contextualize_knowledge}). 

\begin{figure*}[h]
\centering
\includegraphics[width=0.8\textwidth]{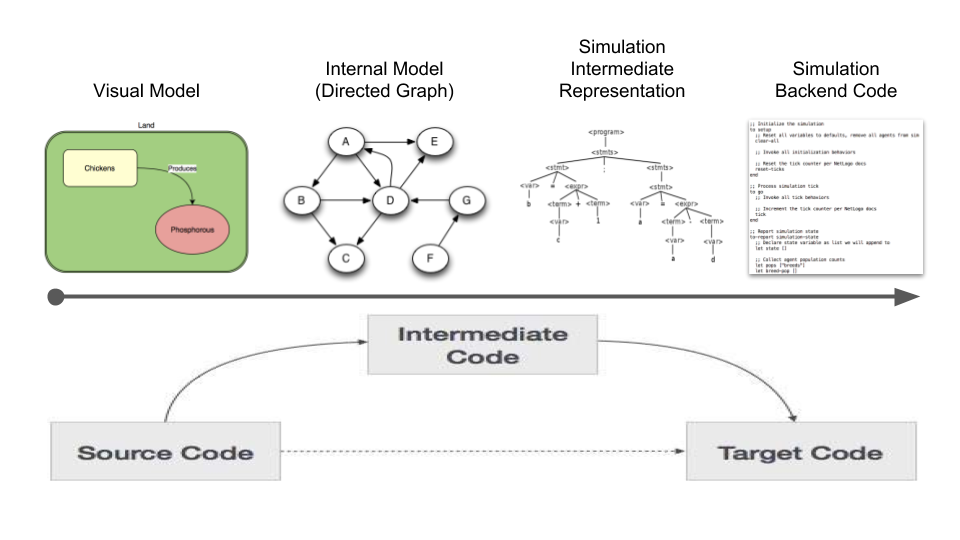}
\caption{Mechanisms of the Visual Conceptual Model to a Simulation.}
\label{fig5}
\end{figure*}

Figure \ref{fig5} illustrates mechanisms for translating the semantics of CMP conceptual models into the semantics of the Netlogo agent-based simulations, which is composed of four steps. First, the compiler extracts conceptual model from visual layout via mxGraph java library. Then it is represented as VeraWeb Domain Model to handle high level simulation concepts. Next, Domain Simulation Builder decomposes high-level domain-specific behaviors and logic into domain independent simulation operations. Finally, the resulting simulation abstract syntax tree (AST) and abstract semantic graph (ASG) compiles abstract simulation into target simulation native constructs such as NetLogo. 

We developed detailed knowledge representations for each component (agent)-component (agent) interaction that will specify how to set up the corresponding simulation. This includes knowledge about the mechanism by which an agent–agent interaction occurs as described in Table \ref{tab2}; for example, in the case of consumption, the consuming organism receives energy and nutrients while the consumed organism perishes. It will also include numerical information relating the various parameters of the mechanism. Some of this information will be specific to the simulation engine so that the AI compiler can in fact automatically generate the corresponding simulation in the given simulation platform.

The components and properties in VERA (in Figure \ref{fig2}) are translated to NetLogo models \cite{wilensky2006thinking}. Biotic components in VERA are translated to Turtles--the primary agents in NetLogo, and their attributes are also translated back to turtles' attributes. For example, carbon parameters, \textit{carbon biomass} (the amount of carbon), \textit{respiratory rate} (loss of carbon biomass), \textit{photosynthesis rate} (gain of carbon biomass), and \textit{assimilation efficiency} (percentage of carbon biomass retained by the consumer) in VERA are calculated to measure \textit{energy} in NetLogo. Spatial parameters, \textit{move velocity} and \textit{move direction}, in VERA are translated to \textit{xcor}, \textit{ycor}, and \textit{heading} in NetLogo (The initial positions are randomly initialized in VERA).  

Figure \ref{fig2}(C) illustrates the time-series graph of the NetLogo simulation results for the conceptual model in Figure \ref{fig2}(A). Note that all four components of the causal model are represented in the simulation: the Light in yellow, Kudzu in green, American Hornbeam in light blue, and Kudzu Bug in purple. Before running a simulation, the user clicks the Reset button to apply input parameters to a new simulation. The user next clicks the Start button to start the time steps of the simulation and clicks the Stop button to pause the simulation.

We should note that VERA is a conceptual modeling tool specifically designed for non-scientists with little scientific training or expertise. Therefore, the core aim of VERA’s agent-based simulation is not to provide accurate scientific predictions, but to give learners the ability to manipulate various variables and to see for themselves how the changes may affect the system. By translating conceptual models into simulations, VERA facilitates the higher-level rapid model revision process. The running of the simulation enables the learner to observe the evolution of the system variables over time, and iterate through the model-simulate-refine loops. In this way, VERA integrates both qualitative reasoning in the conceptual model and quantitative reasoning in the simulation reasoning on one hand, and explanatory reasoning and predictive reasoning. 

\section{Conclusions}
Modeling requires domain knowledge, e.g., relationships between variables describing the system being modeled. The research question in this work is how might we provide access to large-scale domain knowledge for learners engaged in constructing qualitative models of ecological systems in a comprehensible manner? The research hypothesis is that contextualizing large-scale domain knowledge can scaffold the task of authoring conceptual models. VERA contextualizes EOL's large scale domain knowledge to support modeling of ecological systems. It also uses Globi’s ontology of ecological interactions to ground VERA’s CMP models. 

The VERA system has been introduced and evaluated in various contexts including college courses \cite{an2018vera} and self-directed learning on the web \cite{an2022understanding}. Additionally, our data analysis indicates promising results on the usefulness of contextualized domain knowledge in building conceptual models. Based on theses results, the contextualized domain knowledge is expected to help novice learners to use both qualitative and quantitative aspects of ecological knowledge to model, analyze, explain, and predict problems in ecological systems.

\section*{Acknowledgments}
This research was supported by the NSF National AI Insititute
for Adult Learning and Online Education (AI-ALOE; Award \#2112532) and the US NSF grant \#1636848 (Big Data Spokes: Collaborative: Using Big Data for Environmental Sustainability: Big Data + AI Technology = Accessible, Usable, Useful Knowledge!). At Georgia Tech, we thank David Joyner and Taylor Hartman for their contributions to MILA-S, and members of the VERA project, especially Robert Bates. This paper was presented at the 35th International Workshop on Qualitative Reasoning, IJCAI\_ECAI Conference, July 23-29, 2022, Messe Wien, Vienna, Austria.

\bibliographystyle{named}
\bibliography{ijcai22}

\end{document}